\documentclass[pre,floatfix,twocolumn, 10pt]{revtex4-1} 
\usepackage{graphicx}
\usepackage{amssymb}
\usepackage{natbib}
\usepackage{dcolumn}
\usepackage{bm}
\usepackage{color}
\usepackage[colorlinks=true, linkcolor=blue, citecolor=blue]{hyperref}
\usepackage{subfigure}
\usepackage{graphicx,psfrag,xspace}
\usepackage{color}
\usepackage{amssymb,amsfonts,amsmath}
\usepackage{setspace}

\begin{document}

\author{E. A. Jagla} 
\affiliation{Comisi\'on Nacional de Energ\'{\i}a At\'omica, Instituto Balseiro (UNCu), and CONICET\\
Centro At\'omico Bariloche, (8400) Bariloche, Argentina}

\title{The Prandtl-Tomlinson model of friction with stochastic driving}

\begin{abstract} 
We consider the classical Prandtl-Tomlinson model of a particle moving on a corrugated potential, pulled by a spring. In the usual situation in which pulling acts at constant velocity $\dot\gamma$, the model displays an average friction force $\sigma$ that relates to $\dot\gamma$ (for small $\dot\gamma)$ as $\dot\gamma\sim (\sigma-\sigma_c)^\beta$, where $\sigma_c$ is a critical friction force. The possible values of $\beta$ are well known in terms of the analytical properties of the corrugated potential.
We study here the situation in which the pulling has, in addition to the constant velocity term, a stochastic term of mechanical origin 
We analytically show how this term modifies the force-velocity dependence close to the critical force, and give the value of $\beta$ in terms of the analytical properties of the corrugation potential and the scaling properties of the stochastic driving, encoded in the value of its Hurst exponent.
\end{abstract}

\maketitle

\section{Introduction}

\begin{figure}
\includegraphics[width=9cm,clip=true]{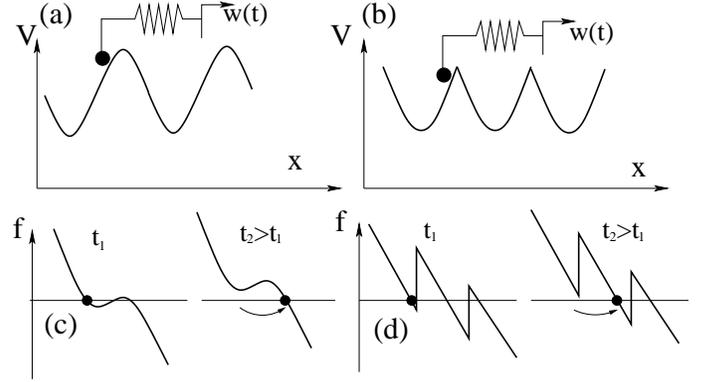}
\caption{Upper panels: Sketch of the PT model, in the two cases studied here of a smooth potential (a) or a piece-wise parabolic potential (b). In (c) and (d) we show the force over the particle close to the instability points, and the transition between one equilibrium point and the next one.
\label{sketch}
}
\end{figure}

The Prandtl-Tomlinson (PT) model \cite{prandtl,tomlinson,popov} describes the movement of a point particle driven by a spring on top of a corrugated potential. It was proposed as a minimum model to understand the origin of a finite kinetic friction force between surfaces, even in the limit of vanishingly small relative velocities. The prototypical equation describing this situation considers a particle with a (one-dimensional) coordinate $x$ that is driven through a spring of stiffness $k_0$ from a point with coordinate $w(t)$, while at the same time feeling the force from an underlying potential $V(x)$ (see a sketch in Fig. \ref{sketch}). 
In the overdamped dynamical regime it reads

\begin{equation}
\eta\dot x=-\frac{dV}{dx}+k_0(w(t)-x)
\label{modelo}
\end{equation}
We will set the bare friction coefficient $\eta$ to 1 from now on.
The average friction force $\sigma$ in the model is defined as the average force that must be applied to the driving point in order to maintain the dynamical evolution. It can be calculated as $\sigma=k_0\overline {(w(t)-x)}$.
The situation of uniform driving corresponds to $w(t)=\dot \gamma t$, where $\dot \gamma$ defines the driving rate. In this case a velocity dependent friction force $\sigma(\dot\gamma)$ exists.
One of the main characteristics of the PT model is the fact that $\sigma$ is finite even if $\dot \gamma \to 0$, [as far as $k_0< \max (-d^2V/dx^2)$], i.e. $\sigma(\dot\gamma\to 0)=\sigma_c\ne 0$. This is due to the fact that in these conditions, there are instability points in the dynamics [Fig. \ref{sketch}(c,d)], in which $x$ jumps by a finite amount and dissipates a finite amount of energy, even in the quasi-static limit $\dot \gamma \to 0$ of the controlling parameter. 

The form of the function $\sigma(\dot \gamma)$ depends fundamentally on the form of $V(x)$. It is usually considered that $V(x)$ is a smooth, oscillating function (a form $\sim \sin(x)$ is typically used as a prototype). In these conditions, for small $\dot\gamma$ it results 
\begin{equation}
\sigma(\dot \gamma)=\sigma_c+C\dot \gamma^{2/3}
\label{tresmedios}
\end{equation}
or in terms of the $\beta$ exponent defined from 
\begin{equation}
\dot\gamma\sim(\sigma-\sigma_c)^\beta
\end{equation}
 we get $\beta=3/2$.
This universal value of $\beta$ is a remarkable feature of the model with a smooth form of $V(x)$. The value of the coefficient $C$ in (\ref{tresmedios}) is not easy to calculate. It was given in some limiting cases in \cite{gnecco}. The overall form of $\sigma(\dot \gamma)$ for arbitrary $\dot \gamma$ 
has not been obtained analytically. We only mention the limiting behavior $\lim_{\dot \gamma\to \infty}(\sigma/\dot\gamma)=1$, that is obtained at large driving velocities, when the potential $V(x)$ becomes irrelevant.
$\beta=3/2$ is originated in the dynamics of the transition from one metastable position to the next, when the energy barrier to escape from the first minimum smoothly vanishes. A qualitatively different situation is obtained when the force is maximum and  discontinuous at the transition point. In this case, the form of $\sigma$ at small $\dot\gamma$ is 
\begin{equation}
\dot\gamma\sim(\sigma-\sigma_c)
\label{tresmedios}
\end{equation}
i.e., the dependence is linear close to the critical point and $\beta=1$.

The PT model has found applications in many areas of pure and applied science. Temperature effect were already considered by Prandtl \cite{prandtl}, and continue to attract attention\cite{muser}.
The model has become crucial in the understanding of experiments using 
atomic force microscopes \cite{meyer}. Extensions to systems of many degrees of freedom are abundant, and the so called Frenkel-Kontorova model \cite{fk1,fk2} is just one case of many possibilities.

Although usually not given this name in this context, the PT model finds also applications in the theory of depinning transitions of elastic manifolds evolving over disordered substrates\cite{fisher,kardar}. In fact, the ``fully connected" version of those models can be represented by a PT model\cite{fisher2}. In that context, the qualitative difference between smooth and parabolic potentials is well known, producing different analytical dependences between strain rate and stress in the system.

In the present paper we want to study the properties of the PT model when there is a stochastic term added
to the uniform driving. The interest in this possibility originates in the study of the plasticity of amorphous materials.\cite{yielding,jagla} In that case, each small part of the system performs a dynamics that corresponds to a particle moving on a corrugated potential while driven by the external deformation, this would correspond to a PT model. In addition, different parts in the system  influence each other due to the coupling mediated by the elasticity of the medium. As the potentials in different spatial positions are uncorrelated, jump
from one potential well to the next occur in an uncorrelated manner across the system, producing a stochastic and cumulative effect on the spatial region under study. This motivates the model we present here. Note that the kind of stochastic noise we are considering is ``mechanical"
in its origin, i.e., the time scale of the noise is the same than the time scale of the external driving. This means that the noise depends directly of the external deformation $\dot\gamma t$ applied to the system. In particular, the noise disappears if $\dot\gamma=0$. This is the situation we will analyze, which is qualitatively different to the case of ``thermal noise", in which the stochastic driving exists independently of the value of $\dot\gamma$. 

The {\em stochastically driven} PT model is thus defined  by a PT model (Eq. (\ref{modelo})) in which the driving term $w(t)$ satisfies

\begin{equation}
\dot w(t)=\dot \gamma+a\sqrt{\dot \gamma} \eta(t)
\label{modelow}
\end{equation}
The $\eta(t)$ is the stochastic term, with the properties $\langle \eta(t)\rangle =0$, $\langle \eta(t)\eta(t')\rangle =\delta(t-t')$. The $\sqrt{\dot\gamma}$ in front of this term is a consequence of its mechanical origin, as mentioned in the previous paragraph. \cite{gamat}

We will consider two different prototypical cases for the potential $V(x)$: the smooth case, with $V(x)=V_0\cos(2\pi x/x_0)$, and a situation with discontinuous forces. For concreteness and in order to get some analytical results, we consider the particular situation in which $V(x)$ is a concatenation of parabolic pieces: $V(x)=V_0 (x/x_0-[x/x_0])^2$, where $[...]$ stands for the nearest integer to the argument. In both cases $x_0$ is the periodicity of the potential.

\section{Parabolic potential}

We concentrate in this section in the case in which the potential force is discontinuous. 
The equations of motion are simply written as 

\begin{eqnarray}
\dot x(t)&=&k_0(w-x)-k(x/x_0-[x/x_0])\\
\dot w(t)&=&\dot \gamma+a\sqrt{\dot \gamma} \eta(t)
\label{eqnarray}
\end{eqnarray}
with $k=2V_0/x_0$.

\begin{figure}
\includegraphics[width=7cm,clip=true]{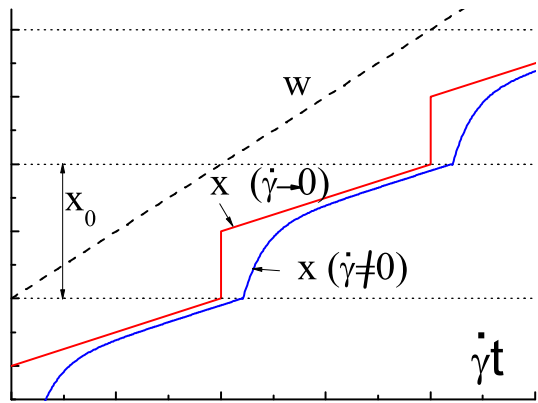}
\caption{Evolution of the coordinates $w$ and $x$ as a function of the applied deformation $\dot\gamma t$, for $\dot\gamma\to 0$, and 
for a finite value of $\dot\gamma$. 
\label{dibujo}
}
\end{figure}

In the absence of stochastic term ($a=0$) $w(t)=\dot\gamma t$ and the dynamics can be solved analytically in all detail. The situation is depicted in Fig. \ref{dibujo}. As the potential is piece-wise parabolic (and then the force piece-wise linear), the dynamics under uniform driving is simply written in term of linear terms and exponential decays. After a bit of algebra matching the solutions at the discontinuity points of the force, the relation between $\sigma$ and $\dot\gamma$ is found to be exactly given by:

\begin{equation}
\sigma(\dot\gamma)=\dot\gamma\frac{k_0(k_0+2k)}{(k_0+k)^2}+\frac{k^2x_0}{2(k_0+k)\tanh{[(k_0+k)x_0/2\dot\gamma}]}
\label{anal}
\end{equation}
For, small $\dot\gamma$, it reduces in linear order to 
\begin{equation}
\sigma\simeq \frac{x_0k^2}{2(k_0+k)}+\dot\gamma\frac{k_0(k_0+2k)}{(k_0+k)^2}
\end{equation}
showing the value of $\sigma_c\equiv  x_0k^2/[2(k_0+k)]$, and the linear increase of $\sigma$ with $\dot\gamma$. For large $\dot \gamma$, Eq. (\ref{anal}) provides

\begin{equation}
\sigma\simeq \dot \gamma+\frac{k^2x_0^2}{12\dot\gamma}+...
\end{equation}

The inclusion of a stochastic part in the driving alters fundamentally the dependence of $\sigma(\dot\gamma)$. We will see that it systematically {\em reduces} the value of $\sigma$, for a given value of $\dot\gamma$. Most remarkably, it changes the analytical dependence of $\dot\gamma$ on $\sigma$ near $\sigma_c$. 

\begin{figure}
\includegraphics[width=8cm,clip=true]{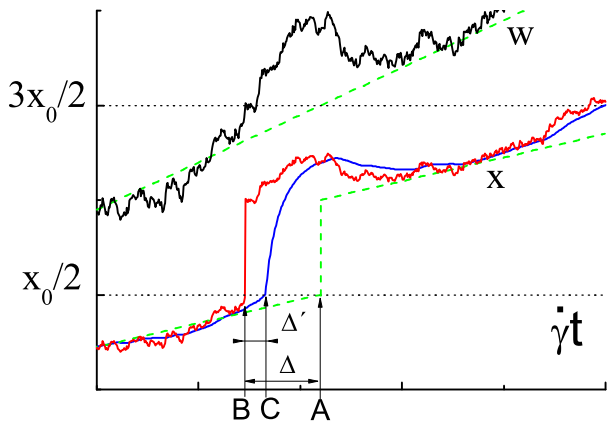}
\caption{Black curve: driving $w$ in the presence of a stochastic component.
Red: evolution of $x$ in the quasistatic limit ($\dot\gamma \to 0$). 
Blue: evolution of $x$ for a finite value of $\dot\gamma$. (Green lines show $w$ and $x$ in the uniform quasistatic case, for comparison. The plots were constructed using $k=k_0$).
\label{dibujo2}
}
\end{figure}
Let us consider first the quasistatic case, $\dot\gamma\to 0$. The situation is depicted in Fig. \ref{dibujo2}. Suppose we are analyzing the movement while $x$ is within the potential well centered at $x=0$. Under quasistatic conditions, the value of $x$ is simple related to $w$ by
\begin{equation}
x=\frac{wk_0}{k+k_0}
\end{equation}
The coordinate $x$ jumps to the next potential well as soon as it reaches the point $x_0/2$. We see in Fig. \ref{dibujo2} that this occurs at point B in the presence of the stochastic term, compared to position A that is the jump in the absence of stochastic term. As the friction force is $k_0\overline {(w-x)}$, there will be a decrease of the kinetic friction of an amount proportional to $\Delta$ with respect to its value in the uniform driving case. The exact value can be obtained in terms of the escape time of a particle diffusing in the interval $[-x_0,x_0]$\cite{redner,gen_rw}. The calculation is a bit lengthy, and we just present here the result for the particular case $k=k_0$. It is obtained that\cite{nota}
\begin{equation}
\sigma_c=\frac{kx_0}{4\tanh(x_0/2a^2)}-\frac{ka^2}{2}
\label{sigmac}
\end{equation}
The next key question is to determine how the kinetic friction increases for finite values of $\dot\gamma$. 
We refer again to Fig. \ref{dibujo2}. When $\dot\gamma$ is finite the time evolution of $x$  (blue curve) becomes smoother compared to the case $\dot\gamma\to 0$. This leads to a delay $\Delta'$ in the jump of $x$ to the new potential well with respect to the $\dot\gamma\to 0$ value, and to an increase of the friction force with respect to $\sigma_c$
of the order of $\Delta'$. 

\begin{figure}
\includegraphics[width=8cm,clip=true]{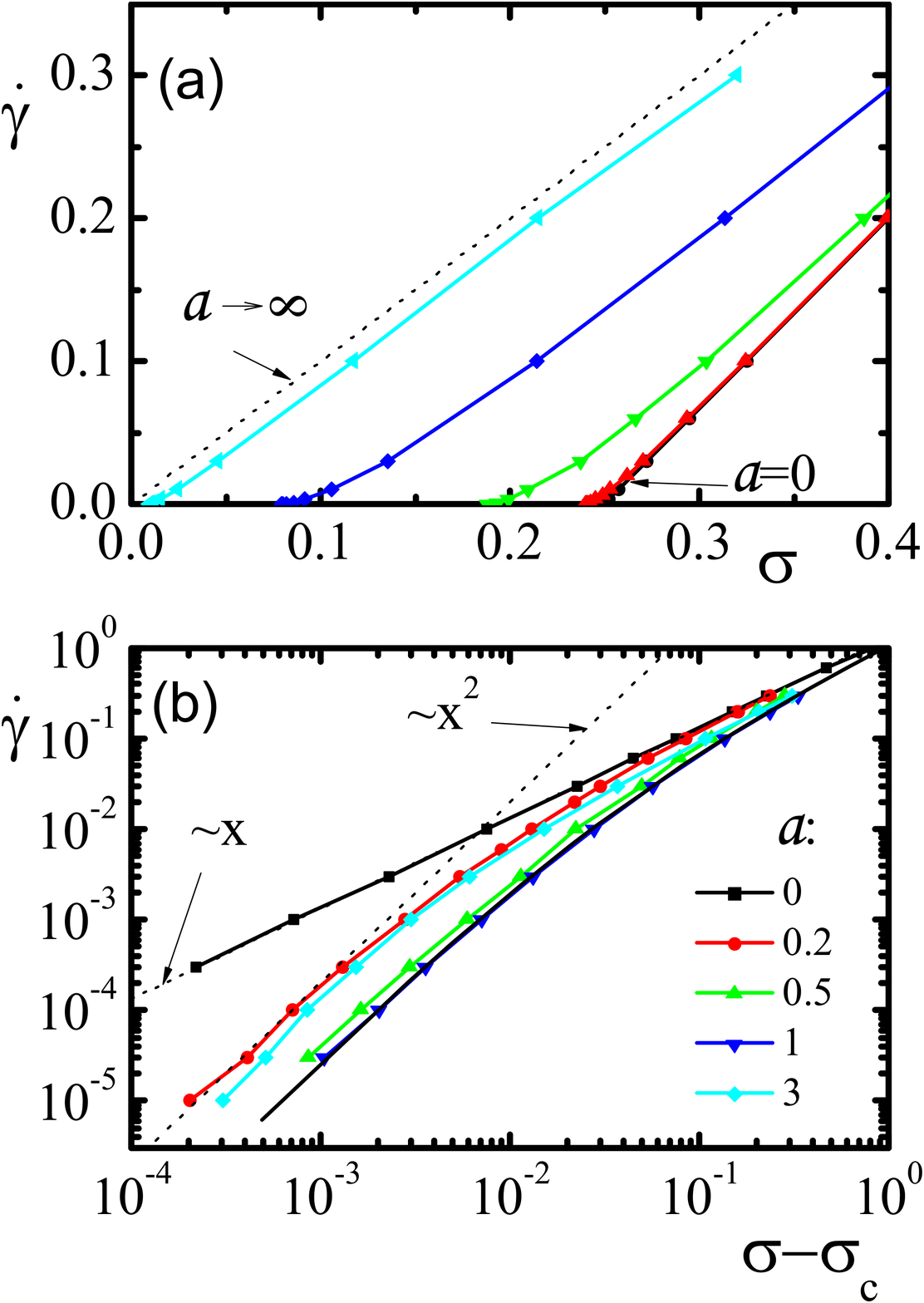}
\caption{Velocity-force curves for parabolic potentials and different intensities of the stochastic component of the driving. Linear (a) and logarithmic scale with respect to the critical force (b). Note that the values of the critical force in each case are taken from the analytical expression (Eq. \ref{sigmac}), and are not fitted. In (b), we show for the $a=1$ case a fitting curve $(\sigma-\sigma_c)=C_1\dot\gamma^{1/2}+C_2\dot\gamma$ (continuous black line).
\label{alfa1h0p5}
}
\end{figure}

If $\dot\gamma$ is sufficiently small, consecutive jumps at $x_0/2$, $3x_0/2$, etc. do not influence each other since $x$ looses memory of the previous jump when it arrives to the next. 
The value of  $\Delta'$ in this case can thus be calculated as $\dot\gamma\tau$, where $\tau$ is time lag of the blue curve with respect to the red one in Fig. \ref{dibujo2} in reaching the coordinate $x_0/2$. Note that before reaching $x_0/2$, the red curve $R$
is simply described by $R=wk_0/(k_0+k)$, whereas the blue one $B$ evolves according to

\begin{equation}
\dot B=(R-B)k_0
\end{equation}
We can have an estimation of $\tau$ by noticing that the quantity $z$ defined as $z\equiv R-B$ evolves as
\begin{equation}
\dot z=-z+\dot\gamma+a\sqrt {\dot\gamma}\eta(t)
\end{equation}
and then its probability distribution $P(z)$ has a mean equal to $\dot\gamma$ and a dispersion $\sim a\sqrt {\dot\gamma}$.
Then when $R$ reaches $x_0/2$ for the first time, $B$ is expected to be at $B\simeq x_0/2-a\sqrt {\dot\gamma}$. From this position, an under a constant drift $\dot\gamma$ it takes a time $\tau\sim a/\sqrt{\dot\gamma}$ to reach the coordinate $x_0/2$.
Coming back to the $\dot\gamma$ dependence of $\sigma-\sigma_c$, we obtain $\sigma-\sigma_c \sim \Delta'= \dot\gamma \tau$, and then
$\dot\gamma\sim (\sigma-\sigma_c)^2$. This is a key result. It shows that the stochastic term changes the value of $\beta$, 
from $\beta=1$ without stochastic term, to $\beta=2$
in the presence of this term. 

This result is confirmed by numerical simulations. In Fig. \ref{alfa1h0p5} we see curves showing the dependence of $\sigma$ on  $\dot\gamma$, for different values of the stochastic term $a$ (in all simulations here and below, we use units such that $k_0=1$, $x_0=1$. In the present case we also use $k=1$). For $a=0$ we see the linear dependence close to $\sigma_c$, whereas for finite $a$ a quadratic regime is observed clearly in the logarithmic plot (Fig. \ref{alfa1h0p5}(b)).
In this respect, we emphasize that the values of $\sigma_c$ used to construct these curves were taken from the analytical expression (Eq. (\ref{sigmac})), and are not a free fitting parameter. Also it can be pointed out that the curves can be fitted rather accurately combining the critical square-root-of-$\dot\gamma$ dependence with a linear term that is expected already in the absence of the stochastic term. An example is provided in Fig. \ref{alfa1h0p5}(b). We also point out that the $\dot\gamma \sim (\sigma-\sigma_c)^2$ dependence is more clearly observed for intermediate value of $a$. In fact, if $a$ is very small its effect is negligible, whereas if it is very large, it produces multiple jumps back and forth among different potential wells, making the effect of the potential to be irrelevant, and the limit $\dot\gamma\sim \sigma$ is approached.

\section{Smooth potentials}

Now we want to extend the results of the previous section to the case in which the potential is smooth. In particular, we will use
the potential $V(x)=V_0\cos(2\pi x/x_0)$. In this case very few results have been obtained analytically even in the case of uniform driving. 
Remarkably, one of the analytical results is the value $\beta=3/2$.
In order to re-obtain this result, and its generalization in the presence of a stochastic term in the driving, we proceed as follows.
Given the equation of motion for $x$
\begin{equation}
\dot x=k\sin(2\pi x/x_0)+(w-x)k_0
\end{equation}
($k=-2\pi V_0/k_0$)
the dependence of $\sigma-\sigma_c$ on the value of $\dot\gamma$ (assumed to be small), is mainly determined by the time the particle spends in passing from a metastable minimum that is vanishing, to the next energy minimum. So we need to consider only the form of the previous equation near the point where the particle is about to destabilize. Generically, we can write 
\begin{eqnarray}
\dot x&=&A|x|^\alpha+w\label{x1}\\
\dot w&=&b\dot \gamma+a\sqrt{\dot\gamma} \eta(t)
\label{x2}
\end{eqnarray}
where we keep a general exponent $\alpha$ for generality. Smooth potentials correspond to $\alpha=2$, the parabolic potential
of the previous section corresponds to $\alpha=1$ \cite{fnote}. Also, we added the redundant parameter $b$, that will be useful in keeping track of different terms that will appear.

In the quasistatic limit $x$ jumps to a new equilibrium position (not contained in Eq. (\ref{x2})) as soon as $\dot \gamma t>0$.
For a finite value of $\dot \gamma$, 
$x$ is lagged with respect to the quasistatic evolution, and it takes an additional time $ \tau$ to arrive at the transition point.
In the absence of stochastic driving ($a=0$) this time can be determined (up to a constant factor) simply by dimensional arguments.
$\tau$ is dependent on the values of $A$, $b$, and $\dot\gamma$ in Eq. (\ref{x2}), and the only combination with the correct units among these quantities is 
\begin{equation}
\tau\sim (b\dot\gamma)^{\frac{1-\alpha}{2\alpha-1}}A^{\frac{-1}{2\alpha-1}}
\label{15}
\end{equation}
This is the order of magnitude of the additional time that the particle spends in a given potential well, before passing to the next. This means an increase in the friction  force (compared to that of the $\dot\gamma\to 0$ case) that is proportional to $\tau$ divided by the total time $t_0$ spent in each potential well, which is $t_0=x_0/(b\dot\gamma)$, i.e. $(\sigma-\sigma_c)\sim \tau b \dot\gamma/x_0$. So we finally obtain the dependence with $\dot\gamma$:
\begin{equation}
\sigma-\sigma_c \sim {\dot\gamma}^{\frac{\alpha}{2\alpha-1}}
\end{equation}
meaning that
\begin{equation}
\beta=2-1/\alpha
\end{equation}
Using $\alpha=2$, it is obtained $\beta={3/2}$
which is the well known result for the PT model with smooth potentials. With $\alpha=1$ we recover the analytical result $\beta=1$ of the previous section.

Now we want to include the effect of the stochastic term ($a\ne 0$ in Eq. \ref{x2}) in order to calculate the exponent of the force-velocity relation.
The value of the lag time $\tau$ should be expressible as a function of $A$, $a$, $b$, and $\dot\gamma$. There are now two independent combinations of these parameters with units of time, that we choose to be
\begin{eqnarray}
t_1&\equiv& a^{\frac{2(1-\alpha)}{\alpha}}b^{\frac{\alpha-1}{\alpha}}A^{-\frac{1}{\alpha}}\label{t1}\\
t_2&\equiv& a^{\frac{(4-6\alpha)}{\alpha}}b^{\frac{4\alpha-2}{\alpha}}A^{-\frac{2}{\alpha}}\dot\gamma\label{t2}
\end{eqnarray}
Dimensional analysis then tells that $\tau$ must be of the form $\tau=t_1 g(t_2/t_1)$ where $g$ is an unknown function. 
In order to be more precise, we need some physically motivated argument.
For instance, in the case in which $\dot \gamma$ is very large, the effect of the stochastic term must be negligible compared with the uniform driving. This means that $\tau$ must be independent of $a$ and we recover the result in Eq. (\ref{15}).
However we are interested in the opposite limit, namely in the case $\dot\gamma \to 0$. Since in this case the stochastic term dominates, one might suppose that the uniform driving is negligible and that $b$ could be set to zero, finding a $\tau$ that is $b$ independent. But this is not correct. 
The reason is that if $b=0$ there is no driving on average, and the time to cross the barrier must diverge. We can obtain the correct scaling when $b\to 0$ by considering not the diverging quantity $\tau$, but the friction force difference $\sigma(\dot\gamma)-\sigma(\dot\gamma\to 0)$. As we already pointed out, this quantity can be estimated as $(\sigma-\sigma_c)\sim \tau b\dot \gamma/x_0$. If $\dot\gamma$ is sufficiently small, the relevance of the uniform driving must be negligible, and this quantity must be $b$-independent. Then we arrive to the conclusion that $\tau\sim 1/b$ for small $\dot\gamma$.

\begin{figure}
\includegraphics[width=8cm,clip=true]{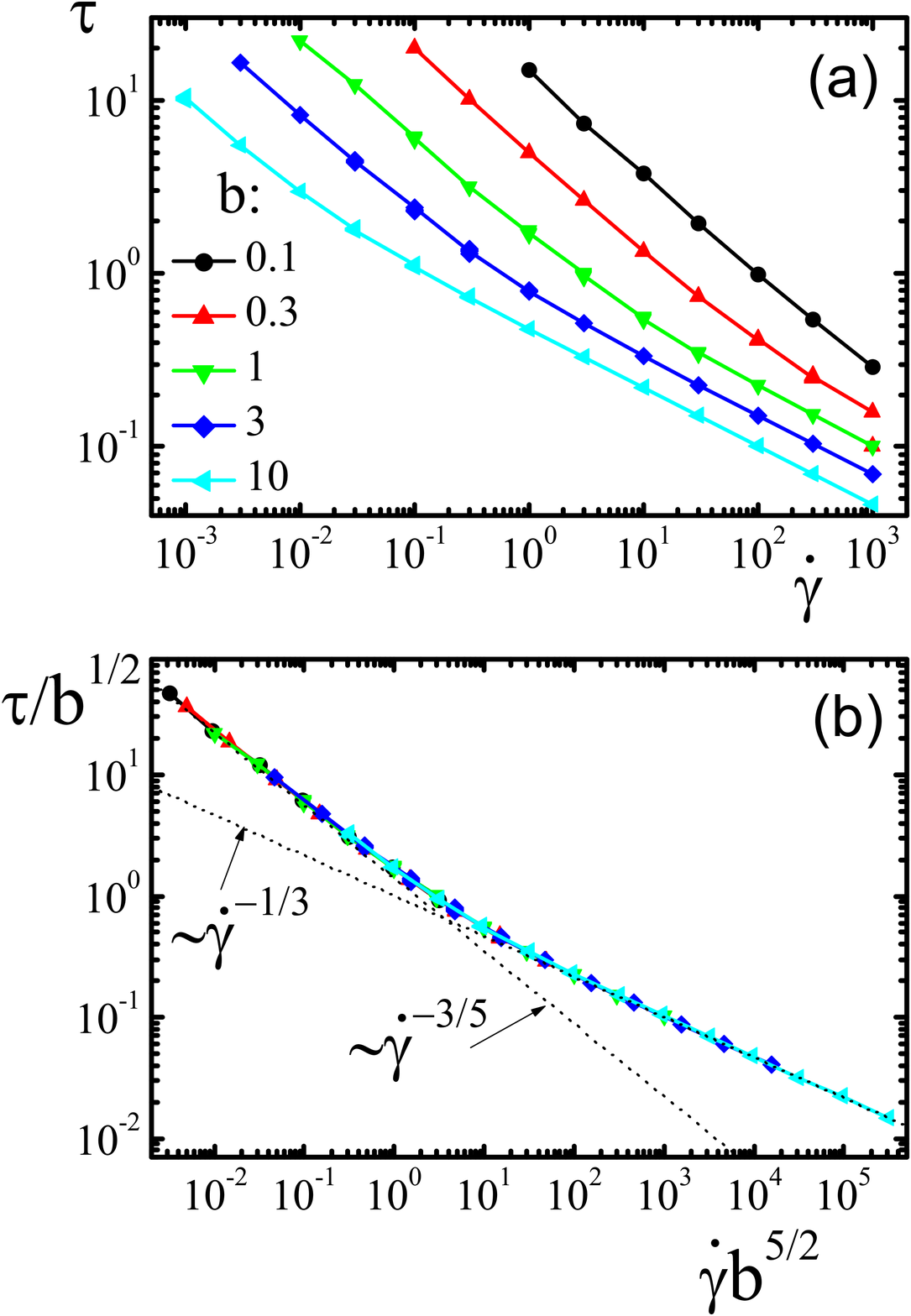}
\caption{(a) Results of numerical simulations of Eqs. (\ref{x1}), (\ref{x2}) with $\alpha=2$. We plot the lag time $\tau$ of $x$ with respect to $w$ in reaching the origin. Parameters used are $a=1$, $A=1$. Curves for different values of $b$ are shown. (b) The data in (a) scaled according to the theoretical expectations. We see the scaling is excellent.
\label{alfa2}
}
\end{figure}

Then constructing $\tau$ using $t_1$ and $t_2$ in such a way that $\tau\sim 1/b$, 
and using the expression for the excess friction force
$(\sigma-\sigma_c) \sim \tau b \dot\gamma/x_0$, we arrive finally to the key result:

\begin{equation}
\sigma-\sigma_c\sim \dot\gamma^{\frac{\alpha}{3\alpha-1}}
\end{equation}
or \cite{non-univ}
\begin{equation}
\beta=3-1/\alpha
\end{equation}

For $\alpha=1$ it reproduces the result of the previous section, namely $\beta=2$.
For the smooth potential case ($\alpha=2$) it gives the value $\beta=5/2$.

Numerical simulations give support to the previous calculations. 
First of all we present results of an implementation of Eqs. (\ref{x1}), (\ref{x2}) in the most important case $\alpha=2$. Numerical integration is made with a standard first order Euler method. To eliminate effects of the initial conditions we set a large and negative initial value of $w$, and follow the time integration determining the times $t_w$ and $t_x$ at which $w$ and $x$ reach zero
for the first time. The process is repeated many times to calculate $\tau=\overline {t_x-t_w}$.
For the parameters chosen ($\alpha=2$, $a=1$, $A=1$) we have $t_1=b^{-1/2}$, $t_2=b^{3}\dot\gamma$. For low $\dot\gamma$ we should get $\tau\sim \dot\gamma ^{-3/5}b^{-1}$. For large $\dot\gamma$ we get (Eq. (\ref{15})) $\tau\sim (b\dot\gamma)^{-1/3}$ instead.
Results are presented in Fig. \ref{alfa2}. We see that the numerical data follow closely the expected dependencies. Note that this confirms in particular the assumed behavior of $\tau\sim 1/b$ for $\dot \gamma\to 0$.

\begin{figure}
\includegraphics[width=8cm,clip=true]{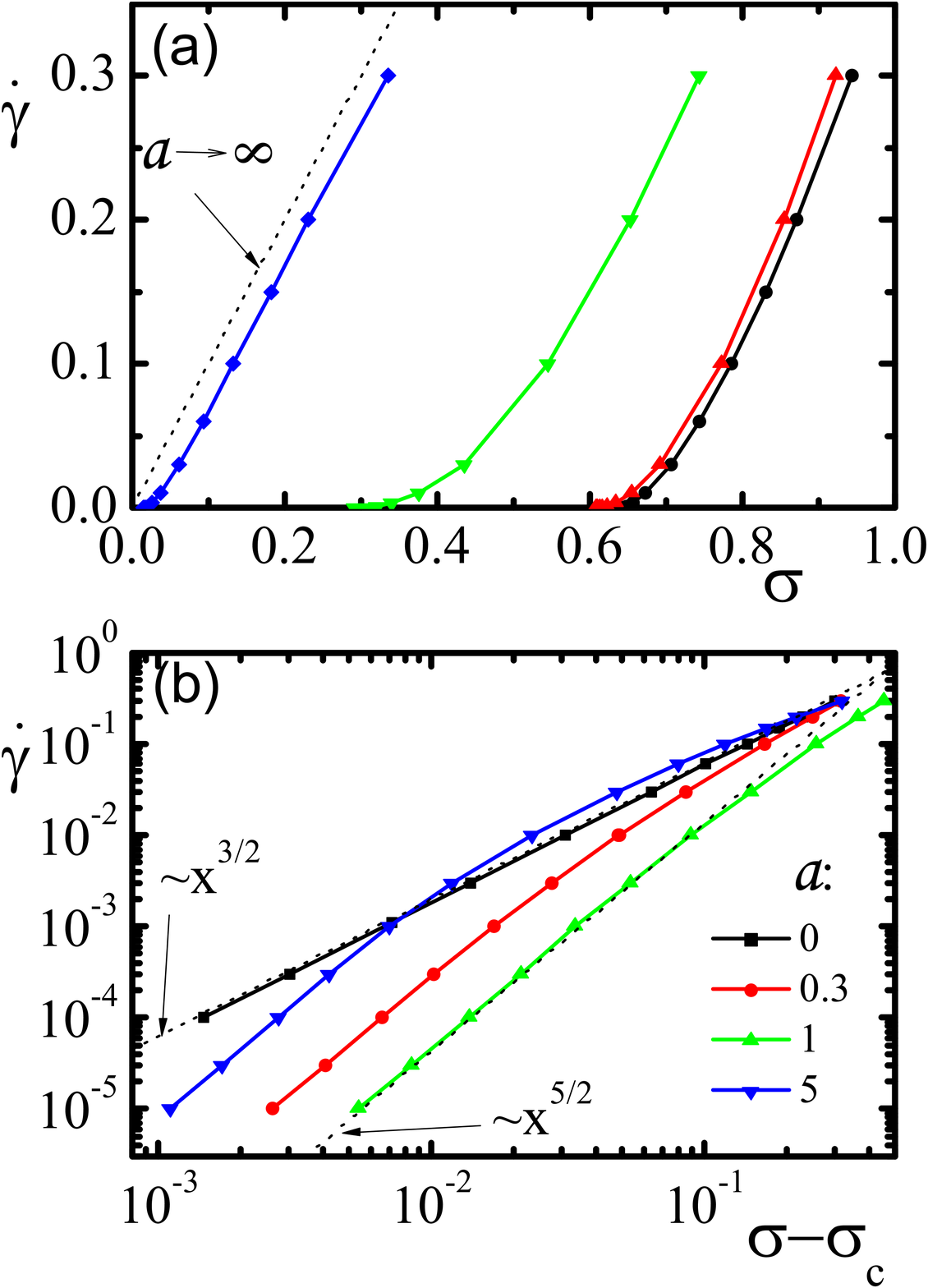}
\caption{Velocity-force curves for smooth potentials and different intensities of the stochastic component of the driving. Linear (a) and log scale with respect to the critical force (b). In this case the value of $\sigma_c$ is adjusted in each case to get a good power law for low values of $\sigma-\sigma_c$. When $a=0$ the value $\beta=3/2$ is observed, whereas $\beta=5/2$ when $a\ne 0$.
\label{alfa2hp5}
}
\end{figure}

Finally we present results of simulation of the PT model in the presence of stochastic driving, for a smooth potential $V(x)=V_0\cos(2\pi x/x_0)$, using $V_0=(2\pi)^{-1}$. The results are presented in Fig. \ref{alfa2hp5}. The data close to the critical point can be very accurately fitted by a law $\dot\gamma=(\sigma-\sigma_c)^\beta$, with the expected value of $\beta$ for this case, which is $\beta=5/2$. Note however that in the present case, the values of $\sigma_c$ were adjusted to get a good power law at small $\dot\gamma$, as the exact values of $\sigma_c$ are not known.

\section{Extension to the case of non-trivially correlated stochastic driving}

An interesting extension of the previous analysis is to consider the case in which the stochastic driving term $w(t)$ is 
more general than the standard random walk already considered.
We will consider a noise term that is characterized by its 
self-similarity properties, such that
\begin{equation}
w_H(\lambda t)\sim \lambda ^H w_H(t)
\end{equation}
$0<H<1$ is called the Hurst exponent\cite{hurst,mandel} of the function $w_H$. 
The case of an usual random walk corresponds to $H=1/2$. 

There are two typical examples of functions $w_H(t)$. One is the case of a fractional Brownian motion.\cite{fbm1,fbm2} This is a case in which the $w_H$ function has non-trivial long time correlations. The second case is obtained as a generalized random walk\cite{gen_rw,infinite_var},  in which the jump function $\eta_H(t)\sim dw_H(t) /dt$ has heavy tails. In concrete, $P(\eta)\sim|\eta|^{-p}$ for large $|\eta|$, 
and by varying $p$ between 2 and 3, functions $w_H$ are obtained  with $H=1/(p-1)$. For any $p>3$, a $w_H$ with $H=1/2$ is obtained. Values of $H$ lower than $1/2$ cannot be obtained by this method.

\begin{figure}
\includegraphics[width=8cm,clip=true]{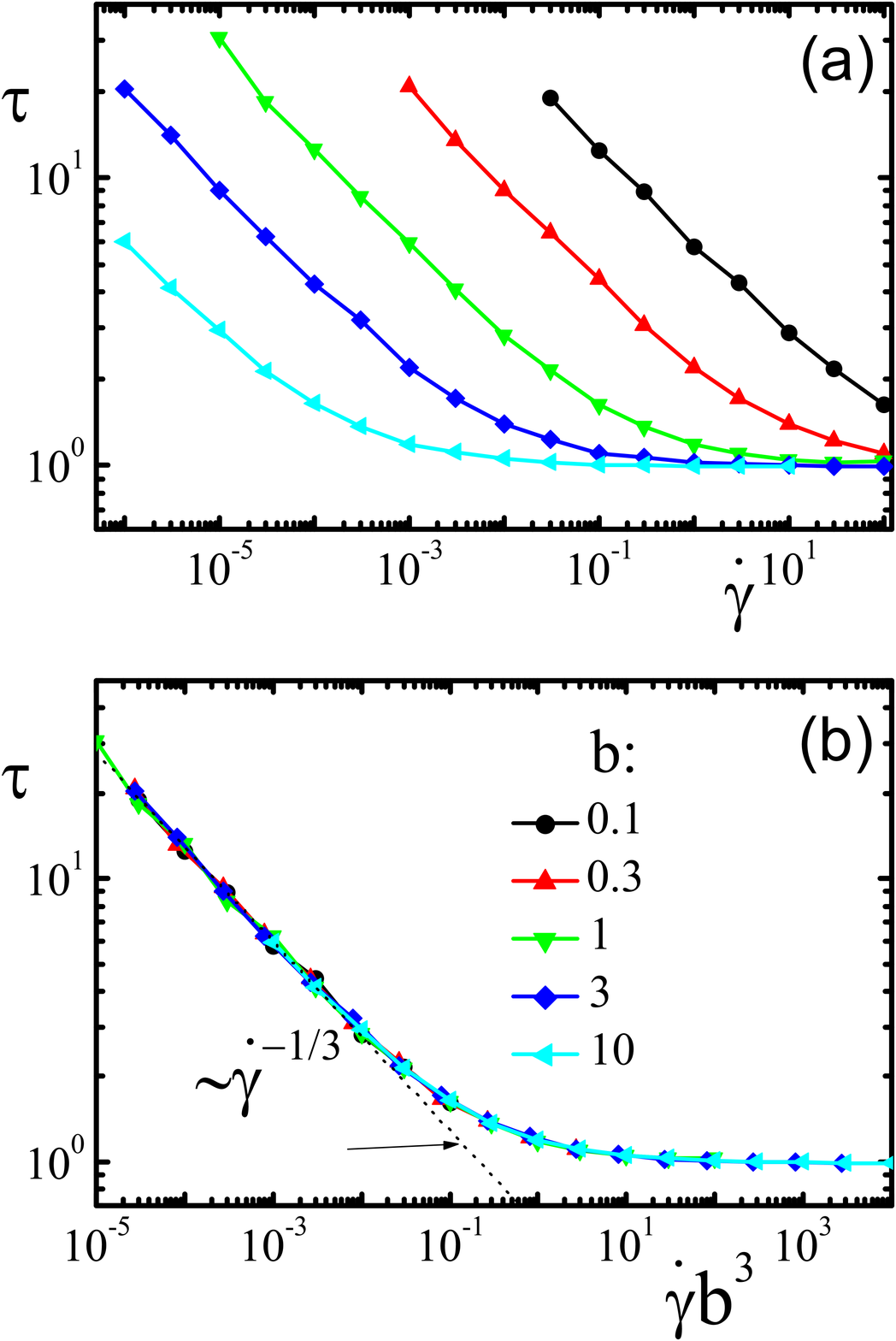}
\caption{(a) Results of numerical simulations of Eqs. (\ref{x1}), (\ref{x2}) with $\alpha=1$, $H=2/3$, $a=1$, and $A=1$. We plot the lag time $\tau$ of $x$ with respect to $w$ in reaching the origin. Curves for different values of $b$ are shown. (b) The data in (a) scaled accordingly to the theoretical predictions.
\label{tau_h2_3}
}
\end{figure}

\begin{figure}
\includegraphics[width=8cm,clip=true]{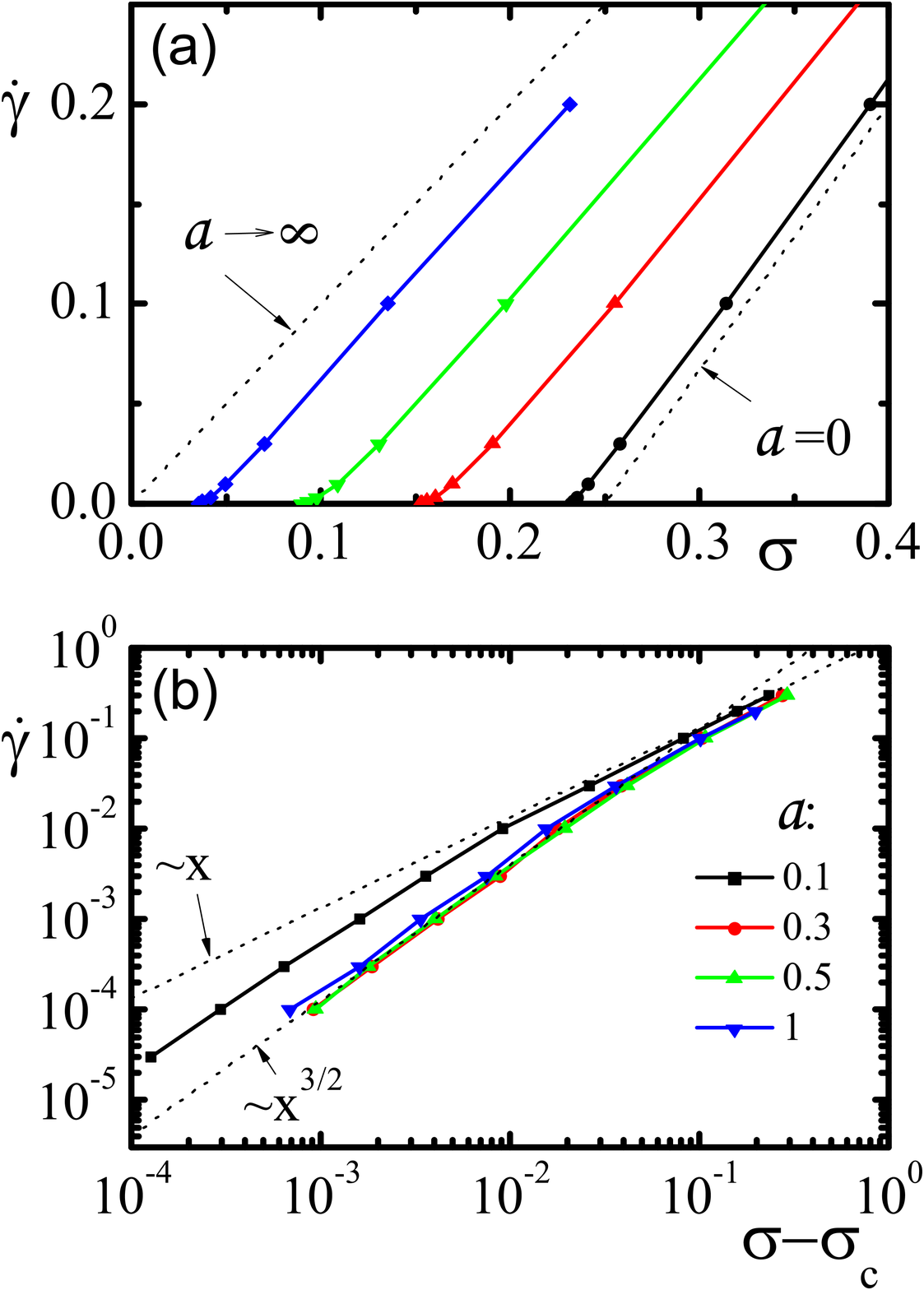}
\caption{Same as Fig. \ref{alfa1h0p5} using a stochastic driving with Hurst parameter $H=2/3$. Linear (a) and log scale with respect to the critical force (b). The value of $\sigma_c$ is adjusted to get a good power law for low values of $\sigma-\sigma_c$. The expected value $\beta=3/2$ is clearly observed.
\label{alfa1h23}
}
\end{figure}

The generic equations of motion close to the instability points that are needed to calculate the time lag $\tau$ are now written as

\begin{eqnarray}
\dot x&=&A|x|^\alpha+w\label{h1}\\
\dot w&=&b\dot \gamma+a\dot \gamma^{H} \eta_H(t)\label{h2}
\end{eqnarray}
where the dimensionality of $\eta_H(t)$ is $[t]^{H-1}$. Proceeding as before, there are two independent quantities with time units that can be constructed, that we choose to be

\begin{eqnarray}
t_1&\equiv& a^{\frac{1-\alpha}{(1-H)\alpha}}b^{\frac{(\alpha-1)H}{\alpha (1-H)}}A^{-\frac{1}{\alpha}}\label{t1h}\\
t_2&\equiv& a^{\frac{2-3\alpha}{\alpha(1-H)}}b^{\frac{2H\alpha-2H+\alpha}{\alpha(1-H)}}A^{-\frac{2}{\alpha}}\dot\gamma\label{t2h}
\end{eqnarray}
Using again the argument that $\tau\sim 1/b$ at low $\dot\gamma$, we can get
from these definitions the scaling of $\tau$ with $\dot\gamma$ as
\begin{equation}
\tau\sim \dot\gamma^{\frac{\alpha-H}{\alpha H+H-\alpha}}
\end{equation}
 and from $\sigma-\sigma_c\sim b\dot\gamma \tau$ we finally obtain

\begin{equation}
\sigma-\sigma_c\sim\dot\gamma^{\frac{\alpha H}{\alpha H +H -\alpha}}
\end{equation}
or simply
\begin{equation}
\beta=\frac 1H -\frac 1\alpha +1
\label{beta-h-alfa}
\end{equation}
Note in particular from here that the difference between $\beta$ values for smooth ($\beta_S$) and parabolic ($\beta_P$) potentials is independent of $H$, and it is $\beta_S-\beta_P=1/2$.

In order to support these findings, we present now some results from numerical simulations. 
They were done using random  walks with heavy tail jump functions $\eta_H$. We choose $\eta$ from a distribution
\begin{equation}
P(\eta)= \frac{1}{2\left (|\eta|+1\right )^{1+\frac 1H}}
\end{equation}
The results presented here were obtained at $H=2/3$, with $\alpha=1$, $a=1$, $A=1$. In this conditions we have
$t_1=1$, $t_2=\dot\gamma b^{3}$. 
For low $\dot\gamma$, and in order to have $\tau\sim 1/b$, we must have $\tau\sim t_1^{4/3}t_2^{-1/3}\sim \dot\gamma^{-1/3}$. 
The results of the numerical simulation are presented in Fig. \ref{tau_h2_3} and fully support the analytical results.

In addition, the simulation of the PT model with piece wise parabolic potentials and a driving with a stochastic component with $H=2/3$ (Fig. \ref{alfa1h23}) displays a value of $\beta$ that is nicely compatible with the expected theoretical value in this case, namely $\beta=1/H=3/2$.

\section{Conclusions}

Summarizing, we have studied the effect of a stochastic term in the driving of a Prandtl-Tomlinson (PT) particle. The case analyzed correspond to that of a ``mechanical noise", in which the time scale of the stochastic term scales with the overall driving rate $\dot\gamma$. In these conditions the system still displays a critical force $\sigma_c$ in its force-velocity dependence. Very generally, this kind of stochastic driving reduces the force for a given value of $\dot\gamma$ compared to the case of uniform driving. Remarkably, it also modifies the critical behavior close to the critical force, namely the value of the $\beta$ exponent in $\dot\gamma \sim (\sigma-\sigma_c)^ \beta$. Our main findings are condensed in expression (\ref{beta-h-alfa}), that gives the value of $\beta$ in terms of the Hurst exponent of the stochastic driving, and the parameter $\alpha$ associated to the analytical properties of the pinning potential. The two most important cases are: $\alpha=2$ for smooth potentials, and $\alpha=1$ when the potential has derivative jumps, as in the case of a concatenation of parabolic pieces. 

The motivation to study the PT model in the presence of a stochastic component in the driving originated in a study of plasticity of materials under shear \cite{jagla}, where the uncorrelated occurrence of plastic events across the system generates  a stochastic contribution to the stress in any given position in the sample. However, due to the multiple scenarios where the PT model has found application, it is expected that the application of the present formalism can have a variety of applications too.


\begin{thebibliography}{37}
\bibitem{gamat}
In this sense, note that Eq. (\ref{modelow}) can be written directly in terms of $y\equiv\dot\gamma t$ as
$dw/dy=1+a\eta(y)$.
\bibitem{nota} We point out that to arrive to this result both forward and backward jumps between different potential wells have been considered.
\bibitem{fnote}For the parabolic case, Eq. (\ref{x1}) is only valid for $x<0$, since as soon as $x>0$ the force becomes finite, of the order of $kx_0$. In the calculation of a transition time then we have to add
and additional contribution $\delta \tau \sim 1$. However this contribution is irrelevant since the value of $\tau$ diverges as $\dot\gamma\to 0$ (see Figs. \ref{alfa2} and \ref{tau_h2_3}).
\bibitem{non-univ}This result was presented previously in \cite{jagla}, but the argument that led to it is questionable.
\end{thebibliography}
\end{document}